\documentclass[showpacs,12pt,preprintnumbers,amsmath,amssymb,aps,nofootinbib]{revtex4}
\usepackage{exscale}                  % correct scaling of math-symbols
\usepackage{amsfonts}
\usepackage{amssymb,amscd}
\usepackage[dvips]{epsfig}                   
\usepackage{array}
\usepackage{wrapfig}
\newcommand{\beqa}{\begin{eqnarray}}
\newcommand{\eeqa}{\end{eqnarray}}
\newcommand{\beq}{\begin{equation}}
\newcommand{\eeq}{\end{equation}}

\newcommand{\cP}{\mathcal{P}}

\def\di{\displaystyle}

\def\bg{\begin{eqnarray}\begin{array}{rcl}\displaystyle}
\def\eg{\end{array} &\di    &\di   \end{eqnarray}}
\def\bm#1{\begin{eqnarray}\begin{array}{#1}\di}
\def\bmo#1{\begin{eqnarray*}\begin{array}{#1}\di}
\def\bml#1#2{\begin{eqnarray}\begin{array}{#1}\label{#2}\di}
\def\bgo{\begin{eqnarray*}\begin{array}{rcl}\displaystyle}
\def\ego{\end{array} &\di    &\di \nonumber  \end{eqnarray*}}

\def\btensor#1#2{\renew\left#1\begin{array}{#2}\di}
\def\brtensor#1#2#3{\ren#3\left#1\begin{array}{#2}}
\def\botensor#1#2{\renew\left#1\begin{array}{#2}}
\def\etensor#1{\end{array}\right#1}

\def\eq#1{(\ref{#1})}
\def\Eq#1{Eq.~(\ref{#1})}

\def\tr{{\rm tr}}

\def\s0#1#2{\mbox{\small{$ \frac{#1}{#2} $}}}
\def\0#1#2{\frac{#1}{#2}}

\def\CO{{\mathcal O}}

\def\CT{{\mathcal T}}

\begin{document}
\preprint{HD-THEP-06-20}
\title{Uniqueness of infrared asymptotics in Landau gauge \\ Yang-Mills 
theory} 
\vspace{-10mm} \author{Christian~S.~Fischer${}^1$ and
  Jan~M.~Pawlowski${}^2$}

\affiliation{${}^1\!\!$ Institut f\"ur Kernphysik,
  Technical University of Darmstadt, Schlossgartenstra\ss e 9, 64289
  Darmstadt,
  Germany\\
  ${}^2\!\!$ 
Institut f\"ur Theoretische Physik, Universit\"at Heidelberg, \\  
Philosophenweg 16, D-69120 Heidelberg, Germany.} %\date{\today}

\begin{abstract}
  We uniquely determine the infrared asymptotics of Green functions in
  Landau gauge Yang-Mills theory. They have to satisfy both,
  Dyson-Schwinger equations and functional renormalisation group
  equations. Then, consistency fixes the relation between the infrared
  power laws of these Green functions. We discuss consequences for the
  interpretation of recent results from lattice QCD. 
\end{abstract}
%05.10.Cc       Renormalization group methods
%11.10.Gh       Renormalization
%11.10.Hi       Renormalization group evolution of parameters
%11.15.-q 	Gauge field theories


\maketitle
%%%%%%%%%%%%%%%%%%%%%%%%%%%%%%%%%%%%%%%%%%%%%%%%%%%%%%%%%%%%%%%%%%%%%%%%%%%%%

\section{Introduction \label{Intro}}

In the past decade much progress has been made in the understanding of
the low energy sector of QCD. This progress has been achieved both
with continuum methods as well as lattice computations. In the
continuum non-perturbative functional methods have been used:
Dyson-Schwinger equations (DSEs) and functional renormalisation group
equations (FRGs). Both frameworks are truly ab initio approaches in
the sense that they can be derived rigorously from the full effective
action of QCD, for reviews see
\cite{Roberts:1994dr,Alkofer:2000wg,Fischer:2006ub,Litim:1998nf%
  ,Berges:2000ew,Polonyi:2001se,Pawlowski:2005xe}.  Although both
frameworks constitute an infinite hierarchy of coupled equations, they
allow for the extraction of scaling laws for Green functions in the
deep infrared \cite{vonSmekal:1997is,Zwanziger:2001kw%
  ,Lerche:2002ep,Pawlowski:2003hq,Alkofer:2004it%,Alkofer:2006gz
}.
These scaling laws are related to important properties of the low
energy limit of QCD, such as confinement and chiral symmetry breaking.

A key building block relevant for the infrared behaviour of QCD are
the ghost and gluon propagators. In Landau gauge the ghost dressing
function gives access to the status of global gauge symmetry: an
infrared diverging ghost unambiguously signals an unbroken symmetry
corresponding to a well-defined global colour charge
\cite{Kugo:1995km}. This is an integral part of the Kugo-Ojima
confinement scenario \cite{Kugo:1979gm}. An infrared vanishing gluon
propagator violates the Osterwalder-Schrader axiom of reflection
positivity \cite{Osterwalder:1973dx}, and transverse gluons do not
belong to the physical asymptotic state space of QCD. Finally, in
Landau gauge one can construct a running coupling with a
renormalisation group invariant combination of ghost and gluon
dressing functions \cite{vonSmekal:1997is}.

In Landau gauge QCD and in terms of correlation functions, the
Kugo-Ojima confinement criterion is expressed as
\begin{eqnarray}\label{eq:KO}
  p^2 \langle A (p) A (-p)\rangle 
  \,\stackrel{p^2 \rightarrow 0}{\longrightarrow}\,0\,,
  \qquad p^2 \langle C(p)\bar C(-p) \rangle 
  \,\stackrel{p^2 \rightarrow 0}{\longrightarrow} \,\infty\,,    
\end{eqnarray}
with the gauge field $A$ and the ghost/anti-ghost fields $C,\bar C$.
An even stronger condition for the gluon propagator has been derived
in a discretised version of Yang-Mills theory, where the infinite
volume/continuum limit can be taken analytically: considerations on 
the impact of the (first) Gribov horizon on dressing functions led to
Zwanziger's horizon condition \cite{Zwanziger:1991gz,Zwanziger:1993qr}
\begin{eqnarray}\label{eq:horizon}
\langle A (p) A (-p)\rangle
  \,\stackrel{p^2 \rightarrow 0}{\longrightarrow}\, 0\,,
\end{eqnarray}
which implies gluon confinement via positivity violation. 

The behaviour \eq{eq:KO} and \eq{eq:horizon} has first been seen in a
DSE-study \cite{vonSmekal:1997is}. This result has been confirmed and
extended within further DSE-computations, {\it e.g.\ }
\cite{Lerche:2002ep,Fischer:2002hn,Fischer:2003rp,Alkofer:2003jj},
stochastic quantisation {\it e.g.\
}\cite{Zwanziger:2001kw,Zwanziger:2003cf} as well as FRG-computations
\cite{Pawlowski:2003hq,Fischer:2004uk,Pawlowski:2004ip,%
  Litim:2004wx}, for related work see also
\cite{Ellwanger:1995qf,Ellwanger:1996wy,Bergerhoff:1997cv,Kato:2004ry}. 
In all these
studies a non-renormalisation theorem for the ghost-gluon vertex
\cite{Taylor:1971ff} is used leading to
\begin{eqnarray}\label{eq:KOT}
  p^2 \langle A (p) A (-p)\rangle\to (p^2)^{2 \kappa}\,,\qquad 
  p^2 \langle C(p)\bar C(-p) 
  \rangle \to (p^2)^{-\kappa}\,,     
\end{eqnarray}  
with $\kappa\in [1/2, 1[$. It has been argued in \cite{Lerche:2002ep}
that \eq{eq:KOT} is the only consistent solution. \Eq{eq:KOT} has been
extended to a self-consistent solution of the (untruncated) tower of
DSEs in continuum Yang-Mills theory \cite{Alkofer:2004it}: 
for proper vertices $Z_{0,\rm as}^{(2n,m)}$ with $n$ ghost, $n$ 
anti-ghost and $m$ gluon legs the infrared asymptotics is given by
\begin{eqnarray}\label{eq:kapnm}
  Z_{0,\rm as}^{(2n,m)}(p^2) \,\,\sim \,\, (p^2)^{\kappa_{(2n,m)}} 
\,\,\sim\,\, (p^2)^{(n-m)\kappa}\, 
\end{eqnarray} 
where $\kappa$ is the exponent of the ghost dressing function
($n=1,m=0$) as defined in \eq{eq:KOT}. 
 
It is interesting to compare the continuum result \eq{eq:KOT} with
results from lattice QCD
\cite{Cucchieri:1997dx,Leinweber:1998uu,Alexandrou:2000ja,%
Langfeld:2001cz,Furui:2004cx,Sternbeck:2005tk,Silva:2005hb,Boucaud:2005ce}.
On the available finite volumes most lattice results for the gluon
propagator are compatible with \eq{eq:KO} and \eq{eq:horizon}. 
Extrapolations towards the infinite volume limit, 
{\it e.g.\ }\cite{Bonnet:2001uh}, seem to agree with an 
infrared finite propagator, {\it i.e.\ } 
\begin{eqnarray} 
  p^2 \langle A(p) A(-p)\rangle \sim (p^2)^1 \,, \label{glue_new}
\end{eqnarray}
see however \cite{Silva:2005hb} for an extrapolation leading to an
infrared vanishing propagator. 
 
The situation is much less clear for the ghost dressing function.
Whereas some simulations give an infrared diverging ghost
\cite{Gattnar:2004bf,Sternbeck:2005tk}, other authors interpret their
results as pointing towards an infrared finite ghost dressing function
\cite{Boucaud:2005ce}, {\it i.e.}
\begin{eqnarray} 
  \langle C(p) \bar C(-p) \rangle \sim
  (p^2)^0 \,. \label{ghost_new} 
\end{eqnarray} 
Clearly, (\ref{glue_new}) and (\ref{ghost_new}) together do not agree
with the continuum result \eq{eq:KOT}. Instead, they have been
proposed as a second possible solution of the continuum DSEs
\cite{Boucaud:2005ce}.

In this work we shall show that the infrared asymptotics of Landau
gauge Yang-Mills is uniquely fixed by DSE and FRG.  We first discuss the
relations between these two sets of equations in the next section.
Then a general infrared analysis of the DSEs for the ghost and gluon
propagators as well as for the ghost-gluon vertex is performed. In the
following section we repeat this analysis within the FRG, and show
that (\ref{glue_new}), (\ref{ghost_new}) cannot survive the infinite
volume/continuum limit. In section~\ref{sec:uniqueness} we show that
self-consistency of DSEs and FRGs enforces the unique solution
(\ref{eq:kapnm}) for the infrared asymptotics of pure Yang-Mills
theory. We briefly discuss the extension of the present analysis to
QCD and the electro-weak sector of the standard model. In our
concluding section we discuss the consequences of this result.

\section
{Functional relations \label{sec:func}} 

A quantum field theory or statistical theory can be defined uniquely
in terms of its renormalised correlation functions. They are generated
by the effective action $\Gamma$, the generating functional of 1PI
Green functions. For the present work we consider pure Yang-Mills
theory with the classical gauge fixed action
\begin{eqnarray}
  S_{\rm cl}=
  \s012 \int \tr\, F^2
  +\s0{1}{2\xi}\int\,(\partial_\mu A^\mu)^2 
  +\int\,\bar C\cdot \partial_\mu D_\mu\cdot C\, . 
\label{eq:sclassical} 
\end{eqnarray} 
in the presence of an additional scale $k$, see
\cite{Pawlowski:2005xe} for a detailed discussion. The propagation is
modified via $k$-dependent terms 
\begin{eqnarray}\label{dSk}
  \Delta S_k=
  \s012 \int \, A^\mu_a\, R_{\mu\nu}^{ab} \, A^\nu_b 
  +\int \, \bar C_a\, R^{ab}\,C_b\,, 
\end{eqnarray}  
where $R_{\mu\nu}^{ab}$ and $ R^{ab}$ are $k$-dependent regulator
functions.  Within the standard choice $k$ is an infrared cut-off
scale, and the functions $R$ cut-off the propagation for momenta
smaller than $k$. Here we also consider more general $R$ that only
have support at about the momentum scale $k$. Such regularisations
allow for a scanning of the momentum behaviour of Green functions. The
regularised effective action $\Gamma_k$ is expanded in gluonic and
ghost vertex functions and reads schematically
\begin{eqnarray}\label{eq:Gk}
  \Gamma_k[\phi]=\sum_{m,n} \0{1}{m! n!^2 } 
  \Gamma_k^{(2n,m)}\, \bar C^n\, C^n\,A^m\,,
\end{eqnarray} 
in an expansion about vanishing fields $\phi=(A,C,\bar C)$. In
\eq{eq:Gk} an integration over momenta and a summation over indices is
understood.  The effective action $\Gamma_k$ satisfies functional
relations such as the quantum equations of motion, the Dyson-Schwinger
equations (DSEs); symmetry relations, the Ward or Slavnov-Taylor
identities (STI); as well as RG or flow equations (FRGs). All these
different equations relate to each other.  Indeed, the Slavnov-Taylor
identities are a projection of the quantum equations of motion,
whereas flow equations can be read as differential DSEs, or DSEs as
integrated flows. Written as a functional relation for the effective
action $\Gamma_k$ and specifying to pure YM theory, the DSE reads,
{\it e.g.\ }\cite{Pawlowski:2005xe}
\begin{eqnarray}\label{eq:DSE} 
  \0{\delta \Gamma_k}{\delta \phi}[\phi]&=&
  \0{\delta S_{\rm cl}}{\delta \phi}[\phi_{\rm op}]\,,
\end{eqnarray} 
where the operators $\phi_{\rm op}$ are defined as 
\begin{eqnarray} 
  \phi_{\rm op}(x) = \int d^4 y\, G_{\phi \phi_i}[\phi](x,y) 
  \0{\delta}{\delta\phi_i(y)}+\phi(x)\,,
\end{eqnarray}
and 
\begin{eqnarray}\label{eq:G}
  G_{\phi_1\phi_2}[\phi]=\left(\0{1}{\Gamma_k^{(2)}[\phi]
      +R_k}\right)_{\phi_1\phi_2}\,
\end{eqnarray} 
is the full field dependent propagator for a propagation from $\phi_1$
to $\phi_2$, with $\phi=(A,C,\bar C)$. The functional derivatives in
\eq{eq:DSE} act on the corresponding fields and generate one loop and
two loop diagrams in full propagators.  The functional DSE \eq{eq:DSE}
relates 1PI vertices, the expansion coefficients of $\Gamma_k$, to a
set of one loop and two loop diagrams with full propagators and full
vertices, but one classical vertex coming from the derivatives of
$S_{\rm cl}$.  We emphasise that the DSE \eq{eq:DSE} only implicitly
depends on the regularisation via the definition of the propagator in
\eq{eq:G}. It has the diagrammatic
representation
%%%%%%%%%%%%%%%%%%%%%%%%%%%%%%%%%%%%%%%%%%%%%%%%%%%%%%%%%%%%%%%%%%%%%%%%
\begin{figure}[h]
\centerline{\epsfig{file=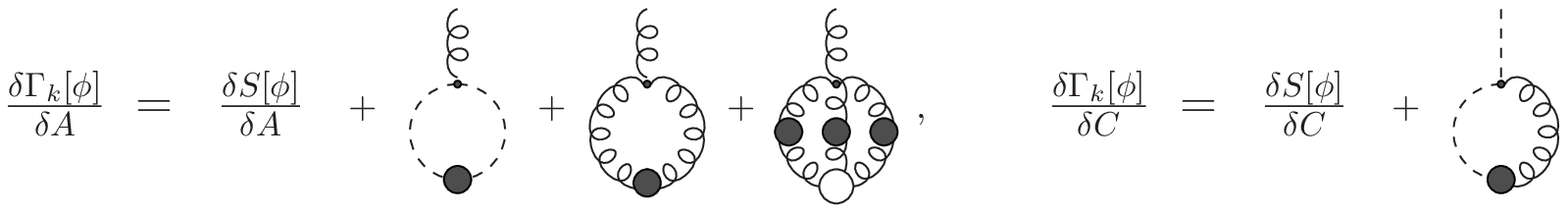,width=16cm}}
\caption{Functional DSE for the effective 
action. Filled circles denote fully dressed field dependent propagators 
\eq{eq:G}. Empty circles denote fully dressed field dependent vertices, 
dots denote field dependent bare vertices.}
\label{fig:funDSE}
\end{figure}
%%%%%%%%%%%%%%%%%%%%%%%%%%%%%%%%%%%%%%%%%%%%%%%%%%%%%%%%%%%%%%%%%%%%%%%%%

Fig.~\ref{fig:funDSE} shows the structure of the functional DSE
\eq{eq:DSE}. The rhs is given in powers of the field-dependent fully
dressed propagator $G_{\phi \phi}[\phi]$, and its derivatives, as well
as the field dependent bare vertices. The momentum scaling of Green
functions is directly related to the scaling of these building blocks.

The flow equation for the effective action reads
\begin{eqnarray}
  \partial_t \Gamma_k[\phi]&=&  
  \frac{1}{2} \int d^4 p \ G_{ab}^{\mu\nu}[\phi](p,p)
  \ {\partial_t} R_{\mu\nu}^{ba}(p) 
  -
  \int d^4 p \ G_{ab}[\phi](p,p)
  \ {\partial_t} R^{ba}(p)\,,
\label{eq:flow}\end{eqnarray} 
where $t=\ln k$. The flow \eq{eq:flow} relates the cut-off scale
derivative of the effective action to one loop diagrams with fully
dressed field-dependent propagators. We can contrast
the diagrammatic representation of the DSE in Fig.~\ref{fig:funDSE}
with that of \eq{eq:flow},
 %%%%%%%%%%%%%%%%%%%%%%%%%%%%%%%%%%%%%%%%%%%%%%%%%%%%%%%%%%%%%%%%%%%%%
\begin{figure}[h]
\centerline{\epsfig{file=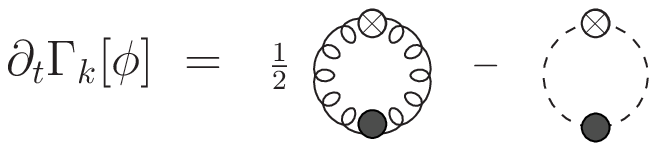,width=6cm}}
\caption{Functional flow for the effective 
action. Filled circles denote fully dressed field dependent propagators 
\eq{eq:G}. Crosses denote the regulator insertion $\partial_t R$. }
\label{fig:funflow}
\end{figure}
%%%%%%%%%%%%%%%%%%%%%%%%%%%%%%%%%%%%%%%%%%%%%%%%%%%%%%%%%%%%%%%%%%%%%%%

Fig.~\ref{fig:funflow} shows the structure of the functional flow
\eq{eq:flow}. The rhs is given by the field-dependent fully dressed
propagator $G_{\phi \phi}[\phi]$ and the regulator insertion
$\partial_t R$. Here, the momentum scaling of Green functions solely
depends on the scaling of $G$ and $\partial_t R$. If we choose the
regulator function $R$ such that it has the RG- and momentum scaling
of the related two point function, the flow is RG-invariant
\cite{Pawlowski:2005xe}, and only depends on full vertices and
propagators, including the regulator term.  The standard use of
\eq{eq:flow} is to take a regulator function $R(p^2)$ which tends
towards a constant in the infrared and decays sufficiently fast in the
ultraviolet, and hence implements an infrared cut-off. In the present
context there is another interesting choice for $R$: let $R$ only have
support at momenta $p$ about the scale $k$, and $R\propto
\Gamma_0^{(2)}$ at momenta $p^2\approx k^2$. Then the regulator term
does not change the theory, $\Gamma_k\simeq \Gamma_0$, and
\eq{eq:flow} only entails the (change of the) momentum dependence of
Green functions of $\Gamma_0$. This provides the resolution of the
momentum dependence at $p^2$ of the full effective action $\Gamma_0$
directly from the flow equation at $k^2=p^2$. \footnote{Such a single
  mode regulator cannot be used to solve the theory by successively
  integrating out degrees of freedom. However, it proves useful for
  studying fixed point solutions \cite{private}.}  We shall detail this
  choice later. It is also worth noting that the relation between DSE
  and FRG is natural in a $N$PI formulation \cite{Pawlowski:2005xe},
  which leads to a mixture of the scaling relations derived from
  \eq{eq:DSE} and \eq{eq:flow}.

In the present work we investigate the leading infrared behaviour of
vertices and propagators constrained by consistency of \eq{eq:DSE} and
\eq{eq:flow}. The crucial ingredient in the related consistency
equations is the fact, that the DSEs derived from \eq{eq:DSE} also
depend on bare or classical vertices whereas the flow equations
derived from \eq{eq:flow} solely depends on full vertices. This allows
us to extract non-trivial information of the theory from a finite set
of vertex DSEs and FRGs that would require a whole infinite tower of
either DSEs or FRGs if restricting the analysis to either of the
functional equations. We analyse the leading infrared behaviour of
\eq{eq:DSE} and \eq{eq:flow} for momenta and cut-offs
\begin{eqnarray}\label{eq:IR} 
  p^2, k^2 \ll \Lambda_{\rm QCD}^2\,.
\end{eqnarray}
To that end we introduce dressing functions $Z_k^{(2n,m)}$ for one
particle irreducible Green functions with $n$ ghost, $n$ anti-ghost
and $m$ gluon legs via
\begin{eqnarray}\label{eq:sym} 
  \Gamma_k^{(2n,m)}(p_1,...,p_{2n+m})=Z_k^{(2n,m)}(p_1,...,p_{2n+m}) 
  \CT^{(2n,m)}(p_1,...,p_{2n+m})\,. 
\end{eqnarray}
The expansion coefficients $\Gamma_k^{(2n,m)}$ of the effective action
have been defined in (\ref{eq:Gk}). The $\CT^{(2n,m)}$ denote the
infrared leading tensor structure of the respective Green function,
and carry their canonical momentum dimension. Then, following the IR
analysis in \cite{Pawlowski:2003hq,Pawlowski:2004ip}, the asymptotic
vertex functions can be expanded about the leading asymptotics at
vanishing cut-off $k=0$:
\begin{eqnarray}\label{eq:vertices} 
  Z_k^{(2n,m)}(p_1,...,p_{2n+m})\simeq Z_{0,\rm as}^{(2n,m)}
  (p_1,...,p_{2n+m})\left(1+
    \delta Z^{(2n,m)}(\hat p_1,...,\hat p_{2n+m})\right)\,,
\end{eqnarray}
where $\hat p_i=p_i/k$, and the asymptotic infrared part $Z_{0,\rm
  as}^{(2n,m)}$ only depends on ratios of monomials and possible
logarithmic dependencies. Inserting the parameterisation
\eq{eq:vertices} into the flow \eq{eq:flow} and solving for $\delta
Z^{(2n,m)}$ one can prove that $\delta Z^{(2n,m)}$ solely depends on
${\hat p}_i$. This suffices to fix the relations between the anomalous
scalings of the vertex functions $Z_{0,\rm as}^{(2n,m)}$ independent
of the $\delta Z^{(2n,m)}$.

For our analysis we only have to know the global scaling behaviour for
the dressing functions $Z_0^{(2n,m)}$, that is, modulo logarithmic scaling,  
\begin{eqnarray}\label{eq:scalings} 
Z_{0,\rm as}^{(2n,m)}(\lambda p_1,...,\lambda p_{2n+m})=
\lambda^{\kappa^{\ }_{2n,m}} Z_{0,\rm as}^{(2n,m)}(p_1,..., p_{2n+m})\,. 
\end{eqnarray}
Specifically interesting for the Kugo-Ojima/Gribov-Zwanziger
confinement scenario are the exponents $\kappa_{0,2}$ and
$\kappa_{2,0}$ of the inverse gluon dressing function $Z^{(0,2)}$,
and inverse ghost dressing function $Z^{(2,0)}$. The horizon conditions  
\eq{eq:KO},\eq{eq:horizon} read  
\begin{eqnarray}\label{eq:kappahorizon} 
\kappa_{2,0}>0\,\qquad \kappa_{0,2}<-1\,.
\end{eqnarray}
We close this section with the remark on the interpretation of the
scaling analysis derived from the combined functional relations
\eq{eq:DSE} and \eq{eq:flow}. In principle, such an analysis produces
the most singular scaling of all diagrams involved and is neither
sensitive to cancellations between different diagrams nor to
cancellations within a given diagram. However, to affect the infrared
behaviour of the Green functions in a consistent way, such
cancellations have to work in the whole tower of DSEs and FRGs and
therefore can only be driven by symmetries. In the present case we
consider such a possibility as highly unlikely. We will come back to
these points at the end of section \ref{sec:uniqueness}.

 \section
 {Infrared analysis of ghost and gluon DSEs \label{DSE}}

\begin{figure}[h]
 \centerline{\epsfig{file=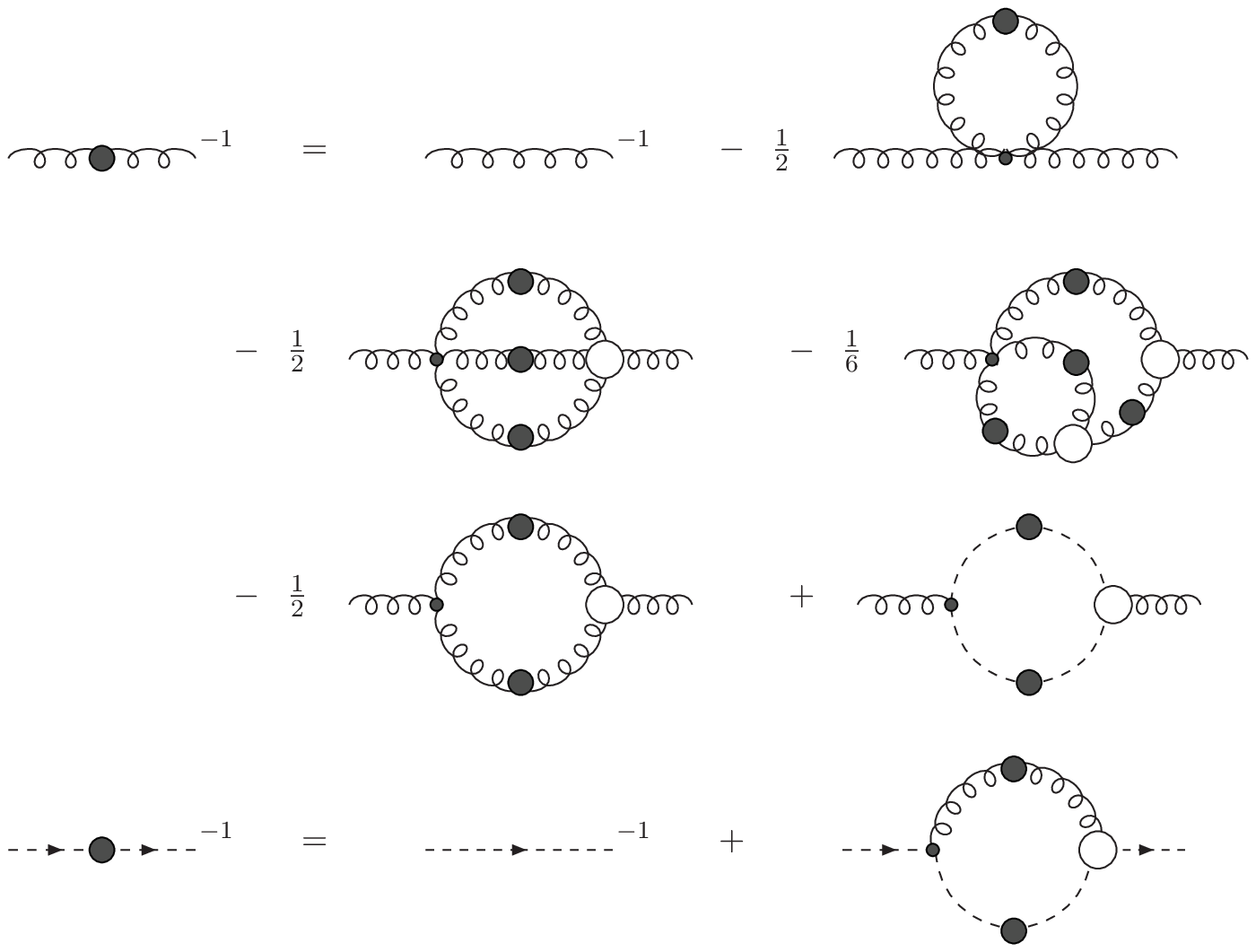,width=12cm}}
 \caption{Dyson-Schwinger equations for the gluon and ghost propagator.
   Filled circles denote dressed propagators and empty circles denote
   dressed vertex functions.}
 \label{DSE-diag}
 \end{figure}

 The Dyson-Schwinger equations for the ghost and gluon propagators are
 given diagrammatically in Fig.~\ref{DSE-diag}. The infrared behaviour
 of these propagators can be analysed as follows 
 \cite{Zwanziger:2001kw,Lerche:2002ep,Alkofer:2004it,Schleifenbaum:2006bq}:
 We choose the external momentum scale $p^2$, according to \eq{eq:IR},
 to be much smaller than
 any other scale, {\it i.e.} $p^2 \ll \Lambda_{\rm QCD}$, where
 $\Lambda_{\rm QCD} \sim \CO (200 \, \mbox{MeV})$ is the
 non-perturbative scale of Yang-Mills theory generated via dimensional
 transmutation. The loop integrals on the right hand side of the DSEs
 are dominated by momentum configurations, where the internal loop
 momentum $q$ is of the same order as the external momentum, {\it i.e.}
 $p^2 \sim q^2$. The reason for this well known behaviour is the
 appearance of at least one propagator in each loop with a denominator
 proportional to $(p-q)^2$.  Thus a self-consistent solution of the
 equations with small external momentum can be obtained by replacing
 all dressing functions inside the loops with their infrared asymptotic
 behaviour.

 %%%%%%%%%%%%%%%%%%%%%%%%%%%%%%%%%%%%%%%%%%%%%%%%%%%%%%%%%%%%%%%%%%%%%%%%%
 %%%%%%%%%%%%%%%%%%%%%%%%%%%%%%%%%%%%%%%%%%%%%%%%%%%%%%%%%%%%%%%%%%%%%%%%%%

 We illustrate this analysis at the example of the ghost DSE. For
 the sake of comparison with the literature we switch to the standard
 DSE notation, where the non-perturbative dressing of the propagators
 is denoted by propagator dressing functions $G(p^2)$ and $Z(p^2)$:
 \begin{eqnarray} 
 \0{1}{Z^{(2,0)}_0(p^2)} = G(p^2) \, , \qquad
 \0{1}{Z^{(0,2)}_0(p^2)} = Z(p^2) \, , 
 \end{eqnarray} 
 and Zwanziger's horizon
 conditions \eq{eq:kappahorizon} read 
 \begin{eqnarray} 
 \lim_{p^2 \rightarrow 0} G(p^2) = \infty \, , \qquad
 \lim_{p^2 \rightarrow 0} \frac{Z(p^2)}{p^2} = 0 \,. \label{horizon} 
 \end{eqnarray} 
 The DSE for the ghost propagator reads is given by
 \begin{eqnarray} 
 \frac{1}{G(p^2)} = \widetilde{Z}_3 - g^2 N_c \int
 \frac{d^4q}{(2 \pi)^4} \frac{G(q^2) Z(l^2)}{p^2 q^2 l^2} \, p \cP(l) q \,
 Z_0^{(2,1)}(p,q)\,, 
 \end{eqnarray} 
 with the momentum routing $l=(q-p)$. The
 abbreviation $p \cP(l) q$ denotes a contraction with the transverse
 momentum tensor $p \cP(l) q = p_\mu P_{\mu \nu}(l) q_\nu$, and
 $Z_0^{(2,1)}(p,q)$ denotes the dressing of the ghost-gluon vertex.
 The ghost renormalisation constant $\widetilde{Z}_3$ absorbs all
 ultraviolet divergencies from the loop integral thus rendering the
 right hand side of the equation UV-finite. This can be made explicit
 within a momentum subtraction scheme. Here $\widetilde{Z}_3$ is
 evaluated at a subtraction point $p^2=\mu^2$, which we choose to be
 $\mu^2=0$. One obtains
 \begin{eqnarray} \widetilde{Z}_3 = \frac{1}{G(0)} + g^2
 N_c \int \frac{d^4q}{(2 \pi)^4} \frac{3}{4}\frac{G(q^2) Z(q^2)}{q^4}
 Z_0^{(2,1)}(0,q)\,.  \end{eqnarray} 
 and subsequently 
 \begin{eqnarray} \frac{1}{G(p^2)}
 &=& \frac{1}{G(0)} - g^2 N_c \int \frac{d^4q}{(2 \pi)^4} \left\{
   \frac{G(q^2) Z(l^2)}{p^2 q^2 l^2}
   \,p \cP(l) q \, Z_0^{(2,1)}(p,q) \right. \nonumber \\
 && \hspace*{4.5cm}+ \left. \frac{3}{4}\frac{G(q^2) Z(q^2)}{q^4}
   Z_0^{(2,1)}(0,q) \right\} \,. \label{ghost-DSE} 
 \end{eqnarray} 
 Now the
 integral is UV-finite and we replace the dressing functions in the
 loop by their infrared expansion in terms of the power laws \begin{eqnarray}
 Z(p^2) \sim (p^2)^{-\kappa_{0,2}} \, ,\qquad G(p^2) \sim
 (p^2)^{-\kappa_{2,0}} \, ,\qquad Z_0^{(2,1)}(p,q) \sim
 (q^2)^{\kappa_{2,1}} \, ,
 \label{power2} 
 \end{eqnarray} 
 The vertex function can be equally well represented by powers of
 $(l^2)$ or, more realistically, by powers of $(p^2+q^2+l^2)$. The
 crucial point is, that after integration all powers of internal loop
 momenta will be transformed into powers of the only external scale
 $p^2$ for dimensional reasons.  This can be seen easily for the
 expansion (\ref{power2}), which leads to integrals that can be
 performed employing 
 \begin{equation}
   \int d^4q \: \frac{(q^2)^a (l^2)^b}{q^2 l^2} = \pi^2 
   \frac{\Gamma(1+a)\Gamma(1+b)\Gamma(-a-b)}
   {\Gamma(1-a)\Gamma(1-b)\Gamma(2+a+b)}\, \, (p^2)^{a+b}\,.
 \label{irintegral}
 \end{equation}
 Plugging (\ref{power2}) into (\ref{ghost-DSE}), performing the
 integration and matching with the left hand side, $1/G(p^2) \sim
 (p^2)^{\kappa_{2,0}}$, we obtain two self-consistent solutions: \begin{eqnarray}
 (p^2)^{\kappa_{2,0}} \sim \left\{
 \begin{array}{l}
 (p^2)^{-\kappa_{0,2}-\kappa_{2,0}+\kappa_{2,1}}\,, \\
 (p^2)^{0}\,. 
 \end{array}
 \right.
 \label{ghost2} 
 \end{eqnarray} 
 In the first case, the loop integral dominates the right hand side,
 and the constant $1/G(0)$ is cancelled by other terms (see
 \cite{Lerche:2002ep} for a detailed discussion). In the second case
 this constant does not vanish and dominates the right hand side of the
 equation in the infrared.  We thus end up with two possible conditions
 \begin{eqnarray} 
 \kappa_{2,0} \sim \left\{
 \begin{array}{l}
 {-\frac{1}{2}\, \kappa_{0,2}+\frac{1}{2}\, \kappa_{2,1}}\,, \\
 {0}\,, 
 \end{array}
 \right.
 \label{kappa-ghost} 
 \end{eqnarray} 
 from the ghost-DSE. Either $G(0)$ is divergent, in accordance with the
 horizon condition (\ref{horizon}), or it is finite as proposed in
 \cite{Boucaud:2005ce} and violates (\ref{horizon}).  
 Strong arguments against the latter 
 possibility have been discussed in \cite{Watson:2001yv,Lerche:2002ep},
 where it has been concluded that $\kappa_{2,0} > 0$. For the sake of
 the argument, however, we will not use this result here but proceed by
 exploring the consequences of both options.

 We would like to stress again that we could have obtained the
 solutions (\ref{kappa-ghost}) without explicitly solving the loop
 integral: since the external momentum $(p^2)$ is the only scale in our
 problem all powers of internal momenta in the loop have to translate
 into powers of external momentum after integration for dimensional
 reasons. Thus by simply counting the infrared exponents of all loop
 propagators and vertices we also arrive at (\ref{kappa-ghost}). 

 We proceed by analysing the DSE for the gluon propagator.
 Schematically we can write this equation as
 \begin{eqnarray} 
   \frac{1}{Z(p^2)} =
   Z_3 + \Pi_{tadpole}(p^2) + \Pi_{sunset}(p^2)
   + \Pi_{squint}(p^2) +
   \Pi_{gluonloop}(p^2) + \Pi_{ghostloop}(p^2)\,, 
 \end{eqnarray} 
 where the dressing loops appear in the same order as in
 Fig.~\ref{DSE-diag}. The static tadpole-term is absorbed in
 the process of renormalisation. We therefore have to analyse the
 infrared behaviour of the four remaining dressing loops.  Counting
 IR-exponents in the loops we arrive at:
 \begin{eqnarray}
   \Pi_{sunset}(p^2) \sim (p^2)^{-3\kappa_{0,2}+ \kappa_{0,4}}
   &\hspace*{5mm}&
   \Pi_{squint}(p^2)    \sim (p^2)^{-4\kappa_{0,2}+2\kappa_{0,3}}  
   \nonumber\\
   \Pi_{gluonloop}(p^2) \sim (p^2)^{-2\kappa_{0,2}+ \kappa_{0,3}}
   &\hspace*{5mm}& \Pi_{ghostloop}(p^2) \sim (p^2)^{-2\kappa_{2,0}+
     \kappa_{2,1}}\,.  
 \end{eqnarray}
 The infrared leading term from the right hand side has to match the
 left hand side $1/Z(p^2) \sim (p^2)^{\kappa_{0,2}}$ of the DSE.  We
 thus obtain the expression
 \begin{eqnarray} 
   \kappa_{0,2} =
   \min\Bigl(0,-3\kappa_{0,2}+ \kappa_{0,4}\,,
   -4\kappa_{0,2}+2\kappa_{0,3}\,, -2\kappa_{0,2}+
   \kappa_{0,3}\,, -2\kappa_{2,0}+ \kappa_{2,1}\Bigr).
   \label{glue1} 
 \end{eqnarray}
 and subsequently 
 \begin{eqnarray} 
   \kappa_{0,2} =
   \min\Bigl(0,\frac{1}{4} \kappa_{0,4}\,, \frac{2}{5}\kappa_{0,3}\,,
   \frac{1}{3}\kappa_{0,3}, -2\kappa_{2,0}+ \kappa_{2,1}\Bigr),
 \label{kappa-glue} 
 \end{eqnarray} 
 as our final condition for the gluon exponent $\kappa_{0,2}$ from the
 gluon-DSE.  In general we expect $\kappa_{0,2} < 0$ according to the
 Kugo-Ojima and horizon conditions (\ref{eq:KO}), (\ref{horizon}) and
 lattice QCD (see {\it e.g.} \cite{Bonnet:2001uh}). It is interesting
 to note that for negative $\kappa_{0,2}$ the contribution from the
 gluon loop, $1/3\kappa_{0,3}$, is never the leading one on the right
 hand side of (\ref{glue1}), since it is always dominated by the
 contribution from the squint diagram, $2/5\kappa_{0,3}$. Thus any
 truncation of the gluon-DSE that assumes a leading gluon loop (see
 {\it e.g.}  \cite{Bloch:2003yu,Aguilar:2004kt,Aguilar:2004sw}) is 
 missing the dominant contribution in the infrared.

 %%%%%%%%%%%%%%%%%%%%%%%%%%%%%%%%%%%%%%%%%%%%%%%%%%%%%%%%%%%%%%%%%%
 \begin{figure}[t]
 \centerline{\epsfig{file=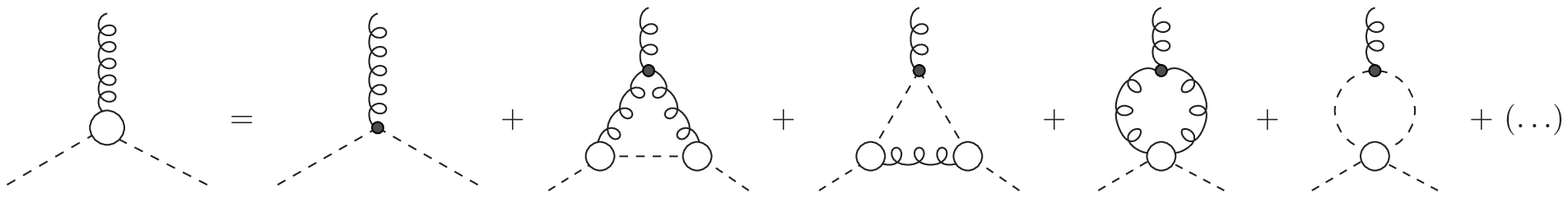,width=15cm}}
 \caption{Dyson-Schwinger equation for the ghost-gluon vertex, derived
   via Eq.~(\ref{eq:DSE}). All internal propagators are taken to be
   fully dressed. The ellipsis denotes two-loop diagrams, which are not
   needed for our analysis.}
 \label{fig:DSE_ghg}
 \end{figure}
 %%%%%%%%%%%%%%%%%%%%%%%%%%%%%%%%%%%%%%%%%%%%%%%%%%%%%%%%%%%%%%%%%%%

 A further crucial ingredient is the DSE for the ghost-gluon vertex.
 One version is derived from the DSE for $\delta\Gamma/\delta A$, and
 is given diagrammatically in Fig.~\ref{fig:DSE_ghg}. Similarly to the
 ghost and gluon propagator DSE we arrive at
 \begin{eqnarray}\label{eq:kappa21DSE} 
   \kappa_{2,1}\leq \min\Bigl(2 \kappa_{2,0}+\kappa_{0,2}\, ,
   \, 2 \kappa_{0,2}+\kappa_{2,0}\,,\, 
   \kappa_{2,2}-2 \kappa_{0,2}\,,\, 
   \kappa_{4,0}-2 \kappa_{2,0}\Bigr)\,.
 \end{eqnarray} 
 The inequality takes into account that the exponents of the two loop 
 diagrams in the vertex-DSE may even be smaller than those of the 
 one-loop diagrams considered in \eq{eq:kappa21DSE}.\footnote{%
   $\kappa_{2,1}$ can be also determined from the DSE for
   $\delta\Gamma/\delta C$ or $\delta\Gamma/\delta \bar C$, see
   Fig.~\ref{fig:funDSE}. However, the present analysis then turns out
   to be more complicated even though two loop terms are absent.} From
 \eq{eq:kappa21DSE} we conclude that
 \begin{eqnarray}\label{eq:kappa21DSE<}
   \kappa_{2,1}\leq 2 \kappa_{2,0}+\kappa_{0,2}\,.
 \end{eqnarray} 
 \Eq{eq:kappa21DSE<} together with the FRG-relation derived in the next
 section suffices to uniquely fix the relations between all $\kappa_{2n,m}$
 in a closed form.

 Based on the conditions (\ref{kappa-ghost}) and (\ref{kappa-glue}) and
 the exact equality in (\ref{eq:kappa21DSE<}) an infrared analysis of
 the DSEs for the three-gluon vertex and the four-gluon vertex has been
 performed in \cite{Alkofer:2004it}. These results have been
 generalised to any Green function with $n$ external ghost, $n$
 anti-ghost and $m$ gluon legs:
 \begin{eqnarray} 
   Z_{0,\rm as}^{(2n,m)}(p^2) \sim (p^2)^{(n-m)\kappa}\,, 
   \label{IRsolution2} 
 \end{eqnarray}
 with $\kappa = \kappa_{2,0} > 0$.  This expression solves
 (\ref{kappa-ghost}), (\ref{kappa-glue}) and any other condition from
 the higher DSEs. In addition it solves the Slavnov-Taylor identities.
 Important aspects of this solution are discussed in detail in
 \cite{Fischer:2006ub}. Two of the characteristic properties of
 (\ref{IRsolution2}) are: (i) contributions from ghost-loops always
 dominate the DSEs and (ii) it leads to IR-fixed points in the running
 couplings from the ghost-gluon ($gh$), three-gluon ($3g$) and
 four-gluon vertex ($4g$). These couplings are defined via
 \begin{subequations}\label{eq:alpha_s}
 \begin{eqnarray}
   \alpha^{gh}(p^2) &=& \frac{g^2}{4 \pi} \, [Z_0^{(2,1)}(p^2)]^2\, 
   G^2(p^2) \, Z(p^2) \,,  
   \label{gh-gl}  \\
   \alpha^{3g}(p^2) &=& \frac{g^2}{4 \pi} \, [Z_0^{(0,3)}(p^2)]^2 \, 
   Z^3(p^2)\,, 
   \label{3g}     \\
   \alpha^{4g}(p^2) &=& \frac{g^2}{4 \pi} \, [Z_0^{(0,4)}(p^2)] \,
   Z^2(p^2)\,, \label{4g} 
 \end{eqnarray}
 \end{subequations} 
 where $Z_0^{-1}=Z_0^{(0,2)}$ and $G^{-1}=Z_0^{(2,0)}$. The vertex
 dressing functions $Z_0^{(2,1)}(p_1,p_2,p_3)$,
 $Z_0^{(0,3)}(p_1,p_2,p_3)$ and $Z_0^{(0,4)}(p_1,p_2,p_3)$ are
 evaluated at the symmetric momentum point $p_1^2=p_2^2=p_3^2=p^2$,
 which make them functions of $p^2$ only. 

 From the tower of DSEs alone it is difficult to prove, that
 (\ref{IRsolution2}) is unique. One way to search for a second possible
 solution would be to assume $\kappa_{2,0}=0$ and $\kappa_{0,2}=-1$
 from the start, corresponding to the behaviour (\ref{glue_new}) and
 (\ref{ghost_new}), as proposed in \cite{Boucaud:2005ce}.  From
 Eqs.~(\ref{kappa-ghost}) and (\ref{kappa-glue}) one obtains
 consistency provided one of the three vertices is strongly divergent:
 \begin{eqnarray}
   \kappa_{2,1}=-1\,,    \quad \mbox{or} \quad 
   \kappa_{0,3}=-5/2\,,  \quad \mbox{or} \quad 
   \kappa_{0,4}=-4 \,. 
 \label{eq:singvertices}
 \end{eqnarray} 

 In the next section we will show that all options
 (\ref{eq:singvertices}) lead to inconsistencies in the functional
 renormalisation group equations. \footnote{Note that the second
   option together with (\ref{3g}) leads to a strongly divergent
   running coupling, $\alpha^{3g} \sim 1/p^2$, which appears to the
   lattice results of \cite{Boucaud:2002fx}.} As discussed in section
 \ref{sec:func}, any solution of the tower of DSEs necessarily has to
 solve the tower of FRGs as well. This provides tight constraints on
 possible solutions, which are in fact sufficient to prove the
 uniqueness of (\ref{IRsolution2}), as we shall see.

 \section
 {Infrared analysis of ghost and gluon flows} \label{sec:FRG}

 Now we repeat the infrared analysis within the FRG framework. We
 restrict ourselves to regulator functions of the form
 \begin{eqnarray}\label{eq:regulators} 
   R_k(p^2)=\Gamma_k^{(2)}(p^2) r(p^2/k^2)\,, 
 \end{eqnarray} 
 where $\Gamma_k^{(2)}$ is the corresponding two point function
 $\Gamma_k^{(2,0)}$ for the ghost, and $\Gamma_k^{(0,2)}$ for the
 gluon.  Regulator functions \eq{eq:regulators} guarantee the
 persistence of the standard RG-scalings in the presence of an IR
 cut-off, and are best-suited for the present studies. 
 Within the parametrisation \eq{eq:vertices} and as a consequence
 of \eq{eq:regulators} propagators $G(p^2)$ take the asymptotic form
 \begin{eqnarray}\label{eq:regprop} 
   G(p^2)=\0{1}{\Gamma_k^{(2)}(p^2)} \0{1}{1+r(\hat p^2)}\simeq 
   \0{1}{\Gamma_{0,\rm as}^{(2)}(p^2)}  \0{1}{1+\delta Z^{(2)}(\hat p)}
   \0{1}{1+r(\hat p^2)}\,.
 \end{eqnarray} 
 \Eq{eq:regprop} can be solely written in terms of $\hat p$ and
 $k$-dependences. Then it reads
 \begin{eqnarray}\label{eq:regprop1} 
   k^{\kappa} 
   \0{1}{\Gamma_{0,\rm as}^{(2)}(\hat p^2)}  \0{1}{1
     +\delta Z^{(2)}(\hat p)}
   \0{1}{1+r(\hat p^2)}\,, 
 \end{eqnarray} 
 where $\kappa$ is either $\kappa_{0,2}$ (gluon) or $\kappa_{2,0}$
 (ghost). The same rescaling can be done with all vertex functions:
 \begin{eqnarray}\label{eq:rescale} 
   \Gamma_k^{(2n,m)}(p_1,...,p_{2n+m})\simeq k^{-\kappa_{2n,m}}
   \Gamma_{0,\rm as}^{(2n,m)}(\hat p_1,...,
   \hat p_{2n+m})(1+\delta Z^{(2n,m)}(\hat p_1,...,\hat p_{2n+m}))\,.
 \end{eqnarray} 

 Another option for the infrared analysis is the choice of a regulator
 function $R_k$ that only has support for momenta at about $k^2$:
 \begin{eqnarray}\label{eq:scan}
   R_k(p^2)=\Gamma_0^{(2)}(p^2) \delta_\epsilon(p^2-k^2)\, 
 \end{eqnarray} 
 where $\delta_\epsilon(x)$ is proportional to a smeared out
 $\delta$-function at $x=0$, the $\delta Z^{(2n,m)}$ only have support
 at momenta $p_i^2\approx k^2$. With regulators \eq{eq:scan}, the
 momentum dependence of $\Gamma_k^{(2n,m)}$ agrees with that of
 $\Gamma_0^{(2n,m)}$.  Only the {\it strength} of the two-point
 function $\Gamma_k^{(2)}\simeq \Gamma^{(2)}_0$ in the momentum window
 $p^2\approx k^2$ is changed. In particular, the infrared power laws at
 $k\neq 0$ agree with those at $k=0$.  Therefore we can directly
 read-off the momentum-dependence at the momentum scale $p^2=k^2$.

 In turn, the general analysis with \eq{eq:rescale} provides additional
 information on the infrared cut-off flow. Within the parameterisation
 \eq{eq:rescale} integrated asymptotic flows for general vertices read
 \begin{eqnarray}\label{eq:asympflow}
   \delta Z^{(2n,m)}(\hat p_1,...,\hat p_{2n +m})\simeq 
   \int_0^k \0{d k'}{k'} \left(\0{1}{\Gamma_{0,\rm as}^{(2n,m)}}
     {\partial_t \Gamma}_k^{(2n,m)}\right)(\hat p_1',...,\hat p_{2n +m}')\,, 
 \end{eqnarray} 
 with possible sub-leading terms. \Eq{eq:asympflow} defines
 consistently renormalised finite DSEs \cite{Pawlowski:2005xe}. In
 contrast to the DSEs \eq{eq:DSE} it solely depends on full vertices
 but also involves an integration over the cut-off scale $k$. The term
 $\partial_t\Gamma^{(2n,m)}$ on the rhs of \eq{eq:asympflow} is derived
 by taking $2n,m$-derivatives of the rhs of the flow \eq{eq:flow},
 leading to a sum of one loop diagrams with dressed vertices and
 dressed propagators.  For vertices with $\kappa_{2n,m}<0$ we have
 \begin{eqnarray}\label{eq:1} 
   \delta Z^{(2n,m)}(0,...,0)=-1\,.
 \end{eqnarray}  
 \Eq{eq:1} simply entails that an infrared cut-off is present and the
 divergent infrared behaviour for $k=0$ is suppressed. Trivially the
 infrared limit \eq{eq:1} only depends on $\hat p$. From \eq{eq:1} we
 derive a relation between the involved $\kappa_{i,j}$ with $i,j\leq
 n+1,m+2$ in the flow of $\Gamma_k^{(2n,m)}$ within an iteration about
 $\delta Z^{(2n,m)}\equiv 0$ on the rhs of \eq{eq:asympflow}.  The
 analysis for $\kappa_{2n,m}\geq 0$ within the integrated flow
 \eq{eq:asympflow} is a bit more involved in one to one correspondence 
 to possible difficulties with bare terms in the DSEs. However, as 
 outlined above we can also directly resolve the momentum behaviour 
 from \eq{eq:flow} with regulators \eq{eq:scan}, where these problems
 are absent. In fact this eliminates the possibility of solely 
 dominating bare terms.

\begin{figure}[h]
 \centerline{\epsfig{file=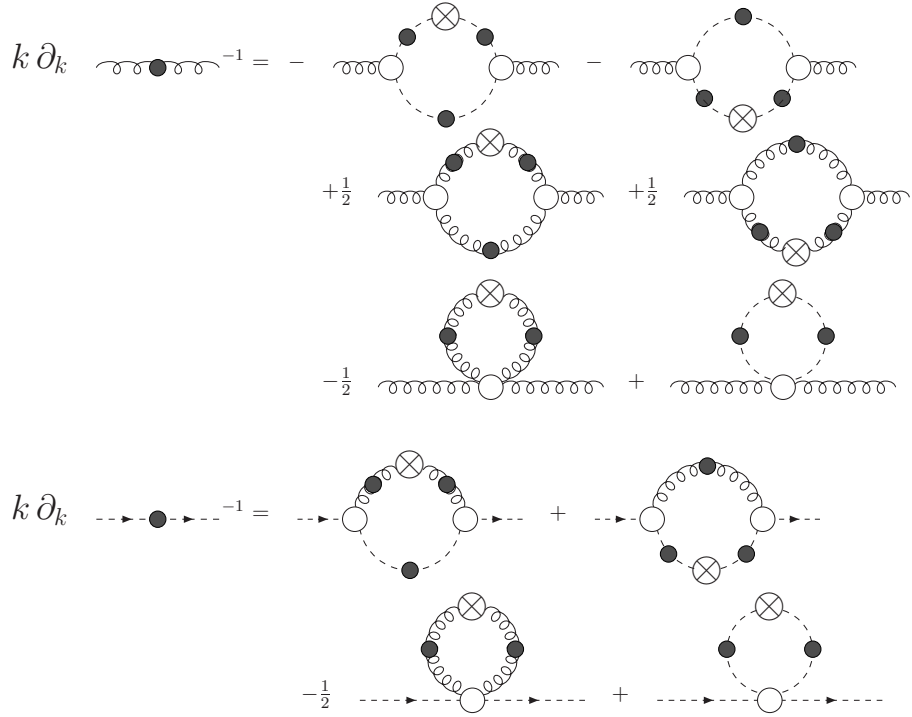,width=12cm}}
 \caption{Functional renormalisation group equations for the gluon and
   ghost propagator. Filled circles denote dressed propagators and
   empty circles denote dressed vertex functions. Crosses indicate
   insertions of the infrared cutoff function.}
 \label{ERGE-diag}
 \end{figure}

 We proceed with the analysis of the propagator FRGs. They can be derived
 from Fig.~\ref{fig:funflow} and are shown diagrammatically in
 Fig.~\ref{ERGE-diag}.
 %%%%%%%%%%%%%%%%%%%%%%%%%%%%%%%%%%%%%%%%%%%%%%%%%%%%%%%%%%%%%%%%%%%%%%%
 %%%%%%%%%%%%%%%%%%%%%%%%%%%%%%%%%%%%%%%%%%%%%%%%%%%%%%%%%%%%%%%%%%%%%%%%%

 We exemplify the analysis at the gluon propagator with
 $\kappa_{0,2}<0$.  As $\delta Z^{(0,2)}$ has to approach $-1$, the
 momentum scaling of the $k$-integral has to precisely cancel that of
 $1/\Gamma_0^{(0,2)}(\hat p\to 0)$. On the rhs of \eq{eq:asympflow} we
 can iterate the full $\Gamma_k^{(2n,m)}$ about those at $k=0$,
 $\Gamma_{0,\rm as}^{(2n,m)}$.  Consequently we can simply sum over the
 $\kappa_{2n,m}$ to identify the leading $\hat p$-behaviour. With the
 regulator \eq{eq:scan} this follows directly, as within this choice we
 have $\Gamma_k^{(2n,m)} \propto \Gamma_{0,\rm as}^{(2n,m)}$ in the flow.
 Then the relations between the $\kappa_{2n,m}$ follow from simple 
 counting of powers of momenta.
 From the flow equation for the gluon propagator we derive the relation
 \begin{eqnarray}\label{eq:gluonkappa} 
   \kappa_{0,2}=\min\Bigl(2\kappa_{2,1}-2 \kappa_{2,0}\,,\, 2\kappa_{0,3}
   -2 \kappa_{0,2}\,,\, \kappa_{0,4}- \kappa_{0,2},\,\kappa_{2,2}
   -\kappa_{2,0}\Bigr) \,, 
 \end{eqnarray}  
 which can be solved for $\kappa_{0,2}$, 
 \begin{eqnarray}\label{eq:gluonkappa1} 
   \kappa_{0,2}=\min\Bigl(2\kappa_{ 2,1}
   -2 \kappa_{2,0}\,,\, \s0{2}{3}\kappa_{0,3}
   \,,\, \s012 \kappa_{0,4},\,\kappa_{2,2}
   -\kappa_{2,0}\Bigr) \,. 
 \end{eqnarray}
 \Eq{eq:gluonkappa1} is already sufficient to rule out all three
 options in (\ref{eq:singvertices}). Indeed, if we insert
 \eq{eq:singvertices} into \eq{eq:gluonkappa1} we arrive at
 \begin{eqnarray} 
   \kappa_{0,2}\leq \left\{
 \begin{array}{l}
   -2\,, \\ -\frac{5}{3}\,, \\ -2\,,
 \end{array}
 \right.
 \label{eq:noncon} 
 \end{eqnarray}
 for the three options. However, \eq{eq:singvertices} goes with
 $\kappa_{0,2}=-1$. The behaviour (\ref{glue_new}) and
 (\ref{ghost_new}), proposed in \cite{Boucaud:2005ce}, is therefore
 ruled out.

 Now we use the combined DSE-FRG analysis to uniquely determine the
 $\kappa$'s without any further input. We shall see that a
 self-consistent solution leads to
 \begin{eqnarray}\label{eq:0} 
   \kappa_{2,1}=0\,.
 \end{eqnarray} 
 To that end we also discuss the derivation of the $\kappa$'s for the
 ghost-propagator and the ghost-gluon vertex. For general regulators
 the infrared analysis for the ghost propagator is intricate and we
 defer the reader to \cite{Pawlowski:2003hq,Pawlowski:2004ip}. With the
 choice \eq{eq:scan} the result follows directly from the flow of the
 ghost propagator. Analogously to \eq{eq:gluonkappa1} we get from
 Fig.~\ref{ERGE-diag}
 \begin{eqnarray}\label{eq:ghostkappa} 
   \kappa_{2,0}=\min\Bigl(\kappa_{2,1}-\012 \kappa_{0,2}\,,\, 
   \kappa_{2,2}-\kappa_{0,2}
   \,, \012 \kappa_{4,0} \Bigr) \,. 
 \end{eqnarray} 
 The FRG relations \eq{eq:gluonkappa1},\eq{eq:ghostkappa} as well as
 the DSE relations \eq{kappa-ghost},\eq{kappa-glue} for $\kappa_{2,0}$
 and $\kappa_{0,2}$ are not closed as they also depend on vertex 
 kappa's.
 The ghost-gluon vertex comprises the crucial information. It is
 protected by non-renormalisation which turns out to be powerful enough
 to fix the whole system completely. Its flow is given by all one loop
 diagrams with regulator insertions and full vertices (up to 5 point
 vertices) with one external gluon line and one ghost and anti-ghost
 line. It reads schematically
 %%%%%%%%%%%%%%%%%%%%%%%%%%%%%%%%%%%%%%%%%%%%%%%%%%%%%%%%%%%%%%%%%%%%%%
 \begin{figure}[h]
 \centerline{\epsfig{file=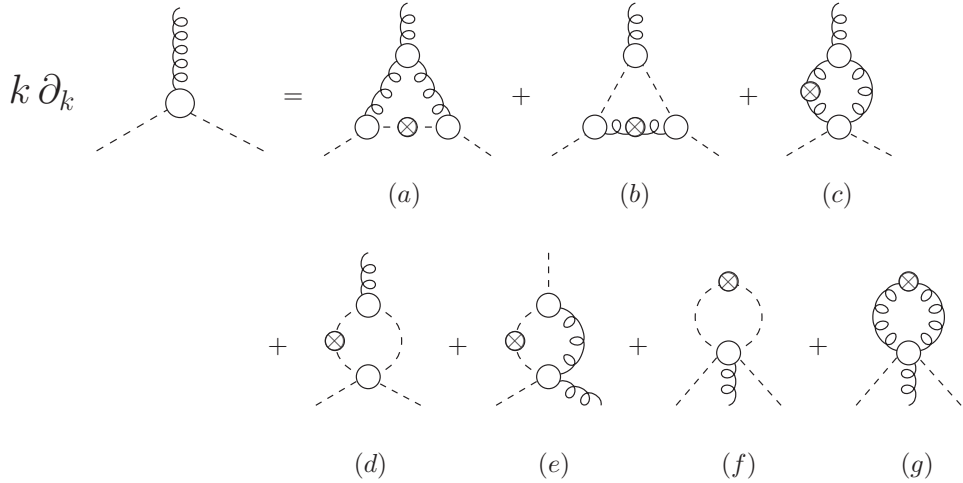,width=13cm}}
 \caption{Functional renormalisation group equations for the ghost
   gluon vertex. All propagators and vertices are fully dressed. Only
   one possible insertion of the infrared cutoff function per diagram
   is shown.}
 \label{fig:FRGghg}
 \end{figure}
 %%%%%%%%%%%%%%%%%%%%%%%%%%%%%%%%%%%%%%%%%%%%%%%%%%%%%%%%%%%%%%%%%%%%

 The infrared analysis of Fig.~\ref{fig:FRGghg} with \eq{eq:asympflow},
 or alternatively employing \eq{eq:scan}, leads to
 \begin{eqnarray}\label{eq:kappa_21}
   \!\!\!\kappa_{2,1}=\min\Bigl(
   2 \kappa_{0,2}+\kappa_{2,0}-\kappa_{0,3}\,,
   \,  \s012 \kappa_{0,2}+\kappa_{2,0}\,,\,
   \kappa_{0,3}+\kappa_{2,2}-2 \kappa_{0,2}\,,\,\kappa_{4,1}
   -\kappa_{2,0}
   \,,\,\kappa_{2,3}-\kappa_{0,2}
   \Bigr),
 \end{eqnarray} 
 from the diagrams $(a),(b),(c),(f),(g)$ respectively. The diagrams
 (d),(e) involve exactly one ghost gluon vertex and the related
 anomalous scaling cancels on both sides of Fig.~\ref{fig:FRGghg}. The
 surviving $\kappa$'s cannot be negative, as this spoils
 self-consistency of the integrated flow. Hence (d),(e) lead to the
 constraints
 \begin{subequations} \label{eq:constraint3C}
   \begin{eqnarray} \label{eq:constraint3C1} \kappa_{4,0}-2
     \kappa_{2,0}&\geq& 0\,,\\\displaystyle
     \kappa_{2,2}-\kappa_{0,2}-\kappa_{2,0}&\geq& 0\,.
     \label{eq:constraint3C2}
 \end{eqnarray}
 \end{subequations} 
 Additionally the non-renormalisation of the ghost-gluon vertex
 \cite{Taylor:1971ff} constrains
 \begin{eqnarray}\label{eq:singghg} 
   \kappa_{2,1}\leq 0\,, 
 \end{eqnarray} 
 as at least one tensor structure of the ghost-gluon vertex has a
 finite dressing. From the second term on the rhs of \eq{eq:kappa_21}
 we extract
 \begin{eqnarray}\label{eq:1kappa_21}
   \kappa_{2,1}\leq \s012 \kappa_{0,2}+\kappa_{2,0}\,. 
 \end{eqnarray}  
 We insert \eq{eq:1kappa_21} in the first term of \eq{eq:gluonkappa}
 and arrive at $ \kappa_{0,2}\leq \kappa_{0,2}$. Consequently the bound
 in \eq{eq:1kappa_21} has to be saturated and
 \begin{eqnarray} \label{eq:kappa21FRG} 
   \kappa_{2,1}=\s012
   \kappa_{0,2}+\kappa_{2,0}\,.
 \end{eqnarray} 
 The DSE-analysis leads to the constraint $\kappa_{2,1}\leq
 \kappa_{0,2}+ 2 \kappa_{2,0}$, \eq{eq:kappa21DSE<}. With
 \eq{eq:kappa21FRG} this turns into $\kappa_{2,1}\leq 2 \kappa_{2,1} $.
 As $\kappa_{2,1}\leq 0$, \eq{eq:singghg}, we arrive at the unique
 solution
 \begin{eqnarray}\label{eq:uniquekappa_21} 
   \kappa_{2,1}=0\,, 
 \end{eqnarray} 
 accompanied by the relation
 \begin{eqnarray}\label{eq:kappa-relation} 
   \kappa_{0,2}=-2 \kappa_{2,0}\,, 
 \end{eqnarray} 
 for the dressing of ghost and gluon propagators in agreement with
 \cite{vonSmekal:1997is}.

\section{Uniqueness of infrared asymptotics} \label{sec:uniqueness}

The analysis of the last two sections for the propagators and the
ghost-gluon vertex can be extended to all $\kappa_{n,m}$.  We first
derive the relations for the purely gluonic three and four point
functions. The flow of three and four gluon vertices is given by all
one loop diagrams with regulator insertions and full vertices (up to 5
and 6 point vertices respectively). For the three gluon vertex this
reads schematically
 %%%%%%%%%%%%%%%%%%%%%%%%%%%%%%%%%%%%%%%%%%%%%%%%%%%%%%%%%%%%%%%%%%%%%%
 \begin{figure}[h]
 \centerline{\epsfig{file=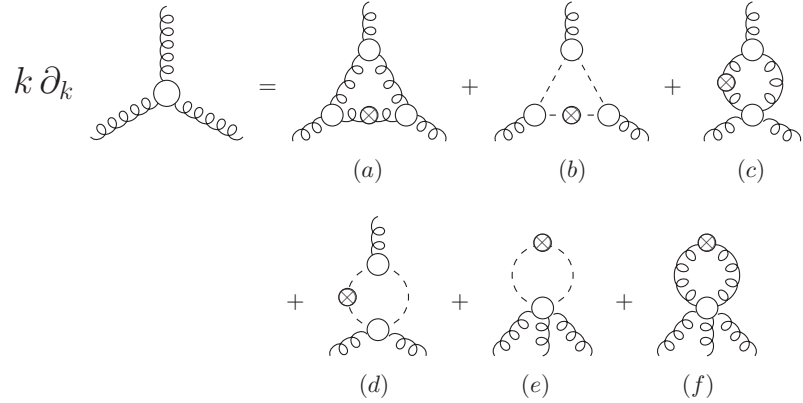,width=11cm}}
 \caption{Functional renormalisation group equations for the three
   gluon vertex. All internal propagators are taken to be fully
   dressed. Only one possible insertion of the infrared cutoff function
   per diagram is shown.}
 \label{fig:FRG3g}
 \end{figure}
 %%%%%%%%%%%%%%%%%%%%%%%%%%%%%%%%%%%%%%%%%%%%%%%%%%%%%%%%%%%%%%%%%%%%%

 Using Fig.~\ref{fig:FRG3g} we arrive at
 \begin{eqnarray}\label{eq:kappa_03}
   \kappa_{0,3}=\min\Bigl(\s032 \kappa_{0,2}\,,\, 3\kappa_{2,1}
   -3\kappa_{2,0}\,,
   \, \kappa_{2,2}+\kappa_{2,1}-2 
   \kappa_{2,0}\,,\, \kappa_{2,3}-\kappa_{2,0}\,,\,\kappa_{0,5}
   -\kappa_{0,2}\Bigr)\,,
 \end{eqnarray} 
 from the diagrams $(a),(b),(d),(e),(f)$. The diagram $(c)$ leads to
 the constraint
 \begin{eqnarray}\label{eq:constraint3} 
   \kappa_{0,4}-2 \kappa_{0,2}\geq 0\,, 
 \end{eqnarray} 
 which already restricts the singular behaviour of the four gluon
 vertex.  We also remark that \eq{eq:kappa_03} puts a simple upper
 bound on $\kappa_{0,3}$, namely
 \begin{eqnarray}\label{eq:boundkappa_03}
   \kappa_{0,3}\leq \s032 \kappa_{0,2}\,. 
 \end{eqnarray} 
 This bound is the natural scaling of the vertex in the presence of a
 fixed point for the coupling constant $\alpha_s$. Note however that a
 priori not all couplings as defined in \eq{eq:alpha_s} run to a fixed
 point.

 The same analysis can be done for the four gluon vertex. Its flow
 reads schematically
 %%%%%%%%%%%%%%%%%%%%%%%%%%%%%%%%%%%%%%%%%%%%%%%%%%%%%%%%%%%%%%%%%%%%%%%%%%%
 \begin{figure}[h]
 \centerline{\epsfig{file=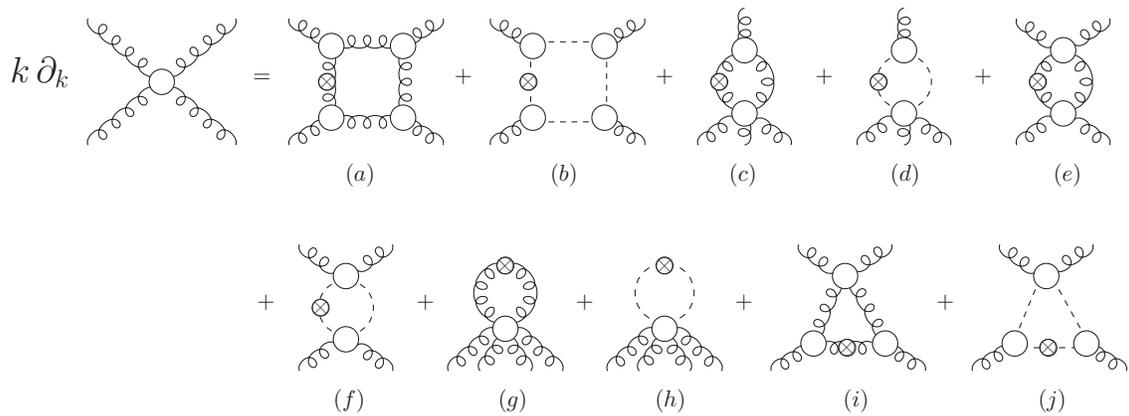,width=15cm}}
 \caption{Functional renormalisation group equations for the four gluon
   vertex. All internal propagators are taken to be fully dressed. Only
   one possible insertion of the infrared cutoff function per diagram
   is shown.}
 \label{fig:FRG4g}
 \end{figure}
 %%%%%%%%%%%%%%%%%%%%%%%%%%%%%%%%%%%%%%%%%%%%%%%%%%%%%%%%%%%%%%%%%%%%%%%%

 Similarly as for the three gluon vertex we derive from
 Fig.~\ref{fig:FRG4g} the anomalous scaling of the four gluon vertex
 \begin{eqnarray}\nonumber 
   \kappa_{0,4}&=&\min\Bigl(4\kappa_{0,3}-4 \kappa_{0,2}\,,\,
   4 \kappa_{2,1}-4 \kappa_{2,0}\, ,\, 
   \kappa_{0,5}+\kappa_{0,3}-2 \kappa_{0,2}\,,\, 
   \kappa_{2,3}+\kappa_{2,1}-2\kappa_{2,0}\,,\, \\ 
   &&  2 \kappa_{0,2}\,,\, 2 \kappa_{2,2}-2 \kappa_{2,0}\,,\,  
   \kappa_{0,6}-\kappa_{0,2}\,,\, \kappa_{2,4}-\kappa_{2,0}\,,\, 
   \kappa_{2,2}+2 \kappa_{2,1}-3  
   \kappa_{2,0}\Bigr)\,,
 \label{eq:kappa_04}\end{eqnarray} 
 from the diagrams $(a),...,(h),(j)$ respectively.  The diagram $(i)$
 leads to the constraint
 \begin{eqnarray}\label{eq:constraint4} 
   2 \kappa_{0,3}-3 \kappa_{0,2}\geq 0\,.
 \end{eqnarray}
 The first term in the second line of \eq{eq:kappa_04} puts a bound on
 $\kappa_{0,4}$,
 \begin{eqnarray}\label{eq:boundkappa04} 
   \kappa_{0,4}\leq  2\kappa_{0,2}\,.
 \end{eqnarray} 
 Together with the constraint \eq{eq:constraint3} this gives the unique
 solution
 \begin{eqnarray}\label{eq:uniquekappa_04} 
   \kappa_{0,4}=  2\kappa_{0,2}\,,
 \end{eqnarray}
 and 
 \begin{eqnarray}\label{eq:uniquekappa_03} 
   \kappa_{0,3}=\s032 \kappa_{0,2}\,. 
 \end{eqnarray}
 Note that the scaling laws
 \eq{eq:uniquekappa_04},\eq{eq:uniquekappa_03} are already generated by
 the diagrams involving ghosts. Indeed, this is valid for all proper
 vertices and leads to the unique solution
 \begin{eqnarray}\label{eq:unique} 
   \kappa_{2n,m}= (n-m)\kappa\,, \qquad {\rm with} 
   \quad \kappa=\kappa_{2,0}\,, 
 \end{eqnarray} 
 which we now prove: first we observe that with \eq{eq:unique} all
 diagrams in the FRG have the same leading infrared asymptotics.
 Let us
 assume for a moment that \eq{eq:unique} is not true for all proper
 vertices. Then at least one vertex has
 \begin{eqnarray}\label{eq:inconsist} 
   \kappa_{2n_0,m_0}< (n_0-m_0)\kappa\,. 
 \end{eqnarray} 
 The vertex $\Gamma^{(2n_0,m_0)}$ occurs in diagrams of FRG for lower
 vertex functions $\Gamma^{(2n,m)}$ with $n_0-n=1$ or $m_0-m=1$.
 Necessarily also these vertices satisfy \eq{eq:inconsist} and
 disagree with \eq{eq:unique}. Within an iteration this enforces that
 {\it all} vertices with $n\leq n_0$ and $m\leq m_0$ satisfy
 \eq{eq:inconsist}, in particular
 $\kappa_{0,4},\kappa_{0,3},\kappa_{0,2},\kappa_{2,1},\kappa_{2,0}$.
 This contradicts the uniqueness of the results
 \eq{eq:uniquekappa_21},
 \eq{eq:kappa-relation},\eq{eq:uniquekappa_04},\eq{eq:uniquekappa_03}
 derived above, and hence proves \eq{eq:unique}.

 For the special case of the propagators this relation has been
 already derived in \cite{vonSmekal:1997is,Lerche:2002ep} with the
 help of additional physical constraints. In \cite{Alkofer:2004it} the
 self-consistency of \eq{eq:unique} for the whole tower of vertex DSEs
 has been shown.  The dominance of ghost loops in the tower of DSEs is
 equivalent to the dominance of the Faddeev-Popov determinant over the
 Yang-Mills action, as proposed in \cite{Zwanziger:2003cf}. In the
 present work we were able to extend these results to a proof of
 uniqueness based on a self-consistency analysis of the quantum
 equations of motion.
 
 The above results hinges on a key structure valid for general
 theories in the presence of a single dynamical scale, and follows
 already from the structure of the functional DSE \eq{eq:DSE},
 Fig.~\ref{fig:funDSE} and FRG \eq{eq:flow}, Fig.~\ref{fig:funflow}:
 any vertex DSE comprises a sum of diagrams proportional to a subset
 of the bare vertices of the theory at hand. Consistency with the FRG,
 which only depends on dressed vertices, requires that in all vertex
 DSEs at least one of these vertices, if dressed, has $\kappa_{\rm
   vertex}=0$. In Landau gauge Yang-Mills this is the ghost-gluon
 vertex.  In general the above criterion leads to more than one vertex
 with $\kappa_{\rm vertex}=0$.
 
 As an example for this general pattern we extend the pure gauge
 theory analysis to a YM-Higgs theory. The functional DSE and FRG can
 be read-off from \eq{eq:DSE},\eq{eq:flow} with $\phi=(A,C,\bar C,h)$
 and an additional Higgs action $S_{\rm higgs}=\s012 \int (D_\mu h)^2+
 V[h]$.  Under the assumption of a single, dynamical mass scale we
 deduce a unique relation for a general vertex function
 $\Gamma^{(2n,m,2l)}$ with $n$ ghost, $n$ anti-ghost, $m$ gluon, $h$
 Higgs lines,
\begin{eqnarray}\label{eq:higgs-relation} 
\kappa_{2n,m,h}=(n-m)\kappa \qquad {\rm with} 
   \qquad \kappa=\kappa_{2,0,0}\,. 
\end{eqnarray} 
In particular it follows that the Higgs propagator has a constant
dressing: $\kappa_h=\kappa_{0,0,2}=0$, as well as the $\phi^4$-coupling
$\kappa_{0,0,4}=0$ required by the presence of vertices with constant 
dressings in all DSEs. Note that $\kappa_{2n,m,l}=0$ includes logarithmic
scaling.  \Eq{eq:higgs-relation} is only valid in the symmetric phase.
In the spontaneously broken phase we expect a massive gauge field
propagator, $\kappa_{0,2,0}=-1$, and massive Higgs propagators,
$\kappa_{h}=-1$.  Furthermore, as a positive $\kappa\geq 0$ for the
ghost signals an unbroken (colour) symmetry, we conclude $\kappa\leq
0$ in agreement with the converse of the Higgs theorem
\cite{Kugo:1979gm}. Due to the additional mass scale some if not all
other vertices may scale canonically, {\it i.e.} $\kappa_{2n,m,h}=0$.
The present infrared analysis then shows consistently
$\kappa_{2n,m,h}\geq 0$, no singular scaling occurs. This is in marked
contrast to massless Yang-Mills theory.

In the case of full QCD, including dynamical quarks, the present
infrared analysis has interesting consequences which shall be
published elsewhere. Finally we discuss the caveat mentioned at the
end of section \ref{sec:func}. From the above analysis it is clear 
non-trivial cancellations always have to occur in an
infinite sub-set of diagrams. It is hard to see which symmetry should
be responsible for such a behaviour, as constraints from gauge
symmetry, {\it i.e.~}STIs, are respected by the solution
\eq{eq:unique}.
 
 \section{Conclusions \label{conclusions}}

 In this work we used Dyson-Schwinger equations (DSEs) and functional
 renormalisation group equations (FRGs) to analyse the infrared
 behaviour of proper vertices of $SU(N_c)$ Yang-Mills theory. We have
 shown that the structure of these functional relations is sufficiently
 different to generate tight constraints for infrared anomalous
 dimensions of these vertices. The caveats of this construction have been
 discussed at the end of sections \ref{sec:func} and \ref{sec:uniqueness}.
 The constraints are powerful enough to
 enforce a unique solution (\ref{eq:kapnm}),(\ref{eq:unique}) for the
 infrared behaviour of proper vertices in the presence of only one
 external scale.  Thus the Kugo-Ojima criterion (\ref{eq:KO}) is
 satisfied ensuring a well-defined global colour charge. A further 
 consequence is
 the fixed point behaviour of the running coupling in the infrared,
 since this behaviour is implied by the solution (\ref{eq:kapnm}) via
 the non-perturbative definitions in Eqs.~(\ref{eq:alpha_s}).
 In turn the proposal (\ref{glue_new}), (\ref{ghost_new}) in
 \cite{Boucaud:2005ce} is excluded. 

 We emphasise that both, the similarities as well as the differences of
 DSEs and FRG were crucial for our results. This structure is certainly
 useful beyond the present investigation, for example if devising 
 truncation schemes. Implicitly this was already used for enhancing the
 respective reliability: the coinciding results for the infrared
 asymptotics from DSE \cite{vonSmekal:1997is,Zwanziger:2001kw%
 ,Lerche:2002ep,Alkofer:2004it%,Alkofer:2006gz
} and FRG
 \cite{Pawlowski:2003hq,Pawlowski:2004ip,Fischer:2004uk} are
 non-trivial as the functional equations are sufficiently different. 

 The present consistency analysis not only uniquely fixes the infrared
 asymptotics but also excludes certain truncation schemes of DSEs and
 FRGs: {\it e.g.} we have shown that truncation schemes of the
 gluon-DSE in Landau gauge Yang-Mills relying on an infrared leading
 behaviour of the gluonic one-loop diagram, as {\it e.g.} assumed in
 \cite{Aguilar:2004kt,Aguilar:2004sw} miss the leading infrared
 behaviour. In addition, truncation schemes that assume all terms in
 the gluon-DSE to be equally leading \cite{Bloch:2003yu} are excluded
 as well.

 It would be desirable to reproduce (\ref{eq:kapnm}) from lattice
 QCD.  To this end one has to address some caveats in comparing
 infrared results from the lattice with those of a continuum
 approach.  Lattice simulations are necessarily performed at finite
 volume and finite lattice spacing and one has to carefully perform
 both an infinite volume and continuum limit extrapolation. These
 procedures are currently under debate
 \cite{Bonnet:2001uh,Silva:2005hb,Cucchieri:2006za%
   ,Boucaud:2006pc,Tok:2005ef}. It is interesting, however, that the
 procedure of \cite{Silva:2005hb} gives $\kappa_{0,2} \approx -1.04$
 in agreement with \eq{eq:KO}.  Unfortunately direct lattice
 calculations in the infrared scaling region $p < 100$ MeV are
 extremely expensive in terms of CPU-time and have not yet been
 performed in four dimensional Yang-Mills theory.  This is different
 in three dimensions, where lattice results are in good agreement with
 the corresponding power-law analysis in the continuum
 \cite{Cucchieri:2003di}.

 Furthermore, gauge fixing is implemented differently. In the
 continuum theory one uses either the Faddeev-Popov method
 \cite{Faddeev:1967fc} or stochastic gauge fixing
 \cite{Zwanziger:2003cf}), whereas on the lattice a gauge fixing
 functional is extremised. Due to the presence of Gribov copies this
 might affect the infrared behaviour of Green functions. These effects
 are currently under investigation in the continuum
 \cite{Zwanziger:2003cf} and on the lattice
 \cite{Cucchieri:1997dx,Alexandrou:2000ja,Silva:2004bv,Furui:2004cx%
   ,Bogolubsky:2005wf,Lokhov:2005ra}.  The effects seem to be much
 stronger for the ghost than for the gluon propagator. This nicely
 corresponds to the fact that lattice and continuum solutions agree
 much better for the gluon than for the ghost.

 The present analysis can be extended to full QCD, and reveals an
 interesting structure. Related results will be published elsewhere.

 %%%%%%%%%%%%%%%%%%%%%%%%%%%%%%%%%%%%%%%%%%%%%%%%%%%%%%%%%%%%%%%%%%%%%%%%%%%%%
 \smallskip
 {\bf Acknowledgements}\\
 CF is grateful to J.~Papavassiliou and J.~Rodriguez-Quintero for
 interesting discussions. We thank R.~Alkofer, H.~Gies, D.F.~Litim and
 L.~von Smekal for a critical reading of the manuscript and useful
 discussions. This work has been supported by the Deutsche
 Forschungsgemeinschaft (DFG) under contract Fi 970/7-1 and GI328/1-2.

 %%%%%%%%%%%%%%%%%%%%%%%%%%%%%%%%%%%%%%%%%%%%%%%%%%%%%%%%%%%%%%%%%%%%%%%%%%%%%

 \end{document}